\documentclass[fleqn,usenatbib]{mnras}
\usepackage[english]{babel}
\usepackage[T1]{fontenc}
\usepackage{ae,aecompl}
\usepackage{fancyhdr}
\usepackage{amsfonts}
\usepackage{amsmath}
\usepackage{amssymb}
\usepackage{multicol}
\usepackage{layout}
\usepackage{graphicx}
\usepackage{multirow}
\usepackage{color}
\usepackage{multirow}
\usepackage{breqn}
\usepackage{everypage}
\usepackage{afterpage}
\usepackage{ifthen}
\usepackage{pgfplots}
\usepackage{times}
\usepackage{natbib}
\usepackage{lipsum}
\def\be{\begin{equation}}
\def\ee{\end{equation}}

\usepackage[utf8]{inputenc}
\usepackage{lmodern}

\definecolor{valecol}{rgb}{0,0.5, 1.}

\newif\ifAMStwofonts
\AMStwofontstrue

\graphicspath{{./fig/}}

	\title{$k-$Dependent Dark Matter}
\author[Arabameri, Davari and Khosravi]{
  Parisa Arabameri$^1$,	Zahra Davari$^2$ and Nima Khosravi$^1$\\ 
	$^1$Department of Physics, Shahid Beheshti University, 1983969411,  Tehran, Iran;\\ $^2$School of Physics,	Institute for Research in Fundamental Sciences (IPM), P.O. Box 19395-5531,  Tehran, Iran\\}

\begin{document}
	\label{firstpage}	
	\maketitle	
\begin{abstract}
	
With the emersion of precise cosmology and the emergence of cosmic tensions, we are faced with the question of whether the simple model of cold dark matter needs to be extended and whether doing so can alleviate the tensions and improve our understanding of the properties of dark matter. In this study, we investigate one of the generalized models of dark matter so that the behavior of this dark matter changes according to the scale of $k$. In large scales (small $k$'s), the dark matter is cold, while it becomes warm for small scales (large $k$'s). This behavior is modeled phenomenologically for two different scenarios. We show that the $S_8$ tension can be alleviated, but the $H_0$ tension becomes milder while not too much.	
\end{abstract}
	

\section{Introduction}\label{intro}
The cold dark matter (CDM) paradigm is an important feature in particle physics and cosmology, assuming cold and collisionless dark matter particles interact only gravitationally. This component is one of the main
bases in the standard $\Lambda$CDM model and it is responsible for about 26\% of the energy density of Universe~\citep{Scott:2018adl, Pan-STARRS1:2017jku}.
 A wide range of cosmological observations from many different epochs and at large and small scales, including CMB missions, BAO data, observations of galaxy clusters, and weak lensing experiments, supported this paradigm. However, the physical nature of DM particles remains unclear and a mystery after decades of research.\\
On the other hand, the CDM paradigm is remarkably successful in many aspects,  especially in explaining the observed properties of large-scale structures (LSS) in the Universe (in the range $\sim$ 1 Gpc down to $\sim$ 10 Mpc); however, it conflicts with observations on galactic and sub-galactic scales ($\leq$1 Mpc). For instance, we can point to:
\begin{itemize}
	\item The "missing satellites problem,"
	which refers to the fact that there is an overestimation of dwarf galaxies by the CDM model than observed in the Universe~\citep{Rubin:1970zza,Rubin:1980zd, Moore:1999nt}.
	\item The "cusp-core problem,"
	which refers to the fact that the CDM model predicts that dark matter halos should have a cuspy density profile at their centers, while observations suggest that they have a more constant density profile~\citep{Gentile:2004tb}.
	\item The "too big to fail problem," which refers to the lack of observation of the most massive halos, which are predicted to be luminous~\citep{Purcell:2012kd}.
\end{itemize} 
The small scale crisis motivated the study of scenarios that predict damped matter fluctuations below a characteristic free-streaming scale through either modification of the primordial power spectrum or non-cold dark matter models, which
modify (suppress) the power spectrum at late times.\\ Furthermore, the recent high-precision cosmological data has shown a statistically significant discrepancy in the estimation of the current values of the Hubble parameter ($ H_0 $) and the fluctuations amplitude of density perturbations at 8 h$^{-1}$Mpc scale ($ \sigma_8 $) between early-time and late-time observations, which poses another challenge to the standard $\Lambda$CDM model. Early universe measurements like CMB Planck collaboration~\citep{Aghanim:2018eyx} estimate
$H_0 \sim(67.0-68.5)$ km/s/Mpc, while late-time
distance ladder measurements like SH0ES and H0LiCOW collaborations report $ H_0=(74.03\pm 1.42) $~\citep{Riess:2019cxk}.\\
The mentioned problems, together with the lack of understanding of the nature, mass, and dynamics of dark matter particles, have sparked several extensions and alternatives to standard dark matter models of particle physics, which are theoretically well-motivated and inspire new search strategies. There are many approaches in order to investigate dark matter, such as warm dark matter (WDM), cannibal Dark Matter~\citep{Buen-Abad:2018mas}, decaying dark matter~\citep{Davari:2022uwd}, dynamical dark matter, fuzzy dark matter and interacting dark matter~\citep{Loeb:2010gj, Archidiacono:2019wdp}.\\
If dark matter particles decouple from the primordial plasma when still relativistic and soon become non-relativistic, the particles are called ``warm dark matter".
These WDM particles would have a smaller free-streaming length than cold dark matter particles, preventing them from clustering on small scales and potentially solving the missing satellite problem.
Furthermore, the WDM particles significantly affect the clustering of matter on large $k$ limit and could flat the inner regions of most galaxies more than the CDM model, reconciling these values with observation and alleviating the core-cusp problem. At the large $k$ limit, DM behaves as WDM as it slightly reduces the DM preferred mass range to a size that includes a moderate initial velocity dispersion and free streaming, sufficient to erase some small scale structures. The suppression in WDM models has a variety of observable implications: abundances of galaxies at high redshift~\citep{Pacucci:2013jfa, Menci:2016eww}, high-redshift gamma-ray bursts (GRBs)~\citep{deSouza:2013wsa}, strong gravitational lensing~\citep{Gilman:2019nap,Hsueh:2019ynk}.
One extension of WDM is to assume that DM comes in two components, a cold one and a warm one, which can be produced via two co-existing mechanisms. These
models are called mixed dark matter (MDM)~\citep{Maccio:2012rjx, Diamanti:2017xfo, Parimbelli:2021mtp}.\\
In this paper, we decided to investigate the case that
dark matter consists of only one component, but its behavior depends on $k$-scale such that in small $k$
it behaves like cold DM, and in large $k$ it shows the properties of warm DM. This scale dependent transition in the behavior can have some motivations in the physics of critical phenomena. The $k$-Dependent dark energy has been studied in~\cite{Farhang:2023hen} based on a phenomenological gravitational phase transition model~\citep{Khosravi:2021csn,Farhang:2020sij}. 

The outline of this paper is as follows: in section~\ref{sec1}, we derive Boltzmann equations governing the evolution at the perturbation level. Then, we implement the related equations in the publicly available numerical code \texttt{CLASS}\footnote{\label{myfootnote0}\url{https://github.com/lesgourg/class_public}}(the Cosmic Linear Anisotropy Solving System)~\citep{Lesgourgues:2011rh} and using the code \texttt{MONTEPYTHON-v3}\footnote{\label{myfootnote}\url{https://github.com/baudren/montepython_public}}~\citep{Audren:2012wb,Brinckmann:2018cvx} to perform a Monte Carlo Markov chain (MCMC) analysis with a Metropolis-Hasting algorithm against the high- CMB TT, TE, EE +low- TT, EE+lensing data from Planck 2018~\citep{Aghanim:2018eyx} in combination with other probes such as the Baryon acoustic oscillations, BAO ( BOSS DR12~\citep{BOSS:2016wmc}, eBOSS Ly-$\alpha$ combined correlations).
\section{Phenomenology of $\lowercase{k}$-Dependent DM model in perturbation level}\label{sec1}	 
In the framework of general relativity, let us consider the flat, homogeneous, and isotropic universe with energy density $\rho(\tau)$ and pressure $P(\tau)$ that is described by the FLRW metric. Using the Einstein equations, we can obtain the following evolution equations for the expansion factor $a(\tau)$.
\begin{eqnarray}
	&&\left(\frac{\dot{a}}{a}\right)^2=\frac{8\pi G}{3}a^2\rho,\\
	&&\frac{d}{d\tau}\left(\frac{\dot{a}}{a}\right)=-\frac{4\pi G}{3}a^2(\rho+3P),
\end{eqnarray} 
where the dots denote derivatives with respect to conformal time, $\tau$. The most convenient way to solve the linearized Einstein equations is in the two gauges in the Fourier space $k$. In the synchronous gauge, the scalar perturbations are characterized by $h(\vec{k},\tau)$ and $\eta(\vec{k},\tau)$. The scalar mode of $h_{ij}$ is given as a Fourier integral 
\begin{equation}
	h_{ij}(\vec{x},\tau)=
	\int d^3k \left(h(\vec{k},\tau){ \hat{k}_i \hat{k}_j}+6\eta(\vec{k},\tau)({\hat{k}_i \hat{k}_j}-\frac{1}{3}\delta_{ij}) \right)e^{i{\vec{k}.\vec{x}}},
\end{equation}
where, h is used to denote the trace of $h_{ij}$ in both the real space and the Fourier space~\citep{Aoyama:2014tga}.\\
The perturbations are characterized by two scalar potentials $\psi(\tau,\vec{x})$ and $\phi(\tau,\vec{x})$ which appear
in the line element as
\begin{equation}
	ds^2= a^2(\tau)\bigg(-(1+2\psi)d\tau^2+(1-2\phi)d\vec{x}^2\bigg),
\end{equation}
and for a perfect fluid of energy density, $\rho$, and pressure, $P$, the energy-momentum tensor has the form 
\begin{equation}
	T^{\mu}_\nu=Pg^\mu_\nu+(\rho+P)u^\mu u_\nu,
\end{equation}
where $u^\mu$ is the four-velocity of the fluid. The perturbed part of energy-momentum conservation equations in $k$-space implies the synchronous gauge as
\begin{eqnarray}\label{eq:wdm}
	&&\dot{\delta}=-3{\cal{H}}(c_s^2-w)\delta-(1+w)(\theta+\frac{\dot{h}}{2}),\\
	&&\dot{\theta}=-{\cal{H}}(1-3c_g^2)\theta+\frac{c_s^2}{1+w}k^2\delta-k^2\sigma,
\end{eqnarray}
and for the conformal Newtonian gauge as
\begin{eqnarray}\label{eq-wdm-n}
	&&\dot{\delta}=-3{\cal{H}}(c_s^2-w)\delta-(1+w)(\theta-3\dot{\phi}),\nonumber\\
	&&\dot{\theta}=-{\cal{H}}(1-3c_g^2)\theta+\frac{c_s^2}{1+w}k^2\delta-k^2\sigma+k^2 \psi.
\end{eqnarray}
The evolution equation for the shear can be obtained as
\begin{equation}\label{eq:sigma}
	\dot{\sigma}=-3\big[\frac{1}{\tau}+{\cal{H}}(\frac{2}{3}-c^2_g-\frac{1}{3}\frac{{\cal{P}}}{p})\big]\sigma+\frac{4}{3}\frac{c^2_{\rm vis}}{1+w}(2\theta+\dot{h}).
\end{equation}
$c_s$ and $c_g$ in the above equations are the effective sound speed and the adiabatic sound speed, respectively. In equation~\ref{eq:sigma}, $c^2_{\rm vis}$ is a new parameter named viscosity speed, and in implementation of \texttt{CLASS}, it is assumed as $c^2_{\rm vis}=3wc^2_g$~\citep{Lesgourgues:2011rh}. The adiabatic sound speed can be expressed as
\begin{equation}
	c^2_g=\frac{\dot{p}}{\dot{\rho}}=w- \frac{\dot{w}}{3{\cal{H}}(1+w)},
\end{equation}
or in another form $c^2_g=\frac{\dot{p}}{\dot{\rho}}=-w \frac{\dot{p}}{p}(\frac{\dot{a}}{a})^{-1}\frac{1}{3(1+w)}
$, that it is stated in~\cite{Tram:2018znz}, $\frac{\dot{p}}{p}=(\frac{\dot{a}}{a})(5-\frac{{\cal{P}}}{p})$. So, the adiabatic sound speed can be rewrote as
\begin{equation}
	c_g^2=\frac{w}{3(1+w)}(5-\frac{{\cal{P}}}{p}),
\end{equation}
here $\cal{P}$ is the pseudo-pressure that for any pressureless species, ${\cal{P}}\simeq p\simeq0$ and for relativistic species we have ${\cal{P}}\simeq p$ since in a higher moment pressure $\frac{{\cal{P}}}{p}\simeq 1$.\\
Obtaining an analytical expression for $c^2_s$ is more complicated since there is no dynamic equation for pressure perturbation, so in ~\cite{Abellan:2021bpx} and ~\cite{Lesgourgues:2011rh}, it is supposed that $c^2_s$ is scale-independent and approximately equal to $c^2_g$. Nevertheless, the full Boltzmann hierarchy calculations show that $c^2_s$ represents a specific $k$-dependence and cannot be obtained with a background quantity such as $c^2_g$, and it increases slightly on the scales $k$. We follow the prescription in~\cite{Abellan:2021bpx} for the synchronous sound speed as
\begin{equation}\label{cs}
	c^2_s(k)=c^2_g\left[1+\frac{1}{5}\sqrt{\frac{k}{k_{fs}}}\right],
\end{equation}
where $k_{fs}=\sqrt{\frac{3}{2}{{\cal{H}}(a)}/c_g(a)}$ is the free-streaming length of the WDM particles.\\
Equations~\ref{eq:wdm}-\ref{eq:sigma} are valid for a single uncoupled fluid or for the net (mass-averaged) $\delta$ and $\theta$
for all fluids. They need to be modified for individual components if the components interact
with each other.\\
The CDM particles can be used to define the synchronous coordinates and therefore have zero peculiar velocities in this gauge. Setting $\theta=\sigma=0$ and $w=\dot{w}=0$ in equation~\ref{eq:wdm} for synchronous gauge lead to 
\begin{equation}
	\dot{\delta}_{\rm CDM}=-\frac{1}{2}\dot{h}.
\end{equation}
\begin{table*}
	\centering
	\caption{The best and mean values and 68\% confidence limit
		(CL) constraints 
		for the free parameters of CDM and two $k$-DM models. They are given using Planck and Planck+Other datasets described in the paper.}
	
\centering 
\resizebox{\textwidth}{!}{\begin{tabular}{|c|c|c|c|c|c|} 
		
		\hline   
		Model &Parameter&\multicolumn{2}{c|}{$Planck$}  &\multicolumn{2}{c|}{$Planck+Other$}  
		\\\cline{3-6}   

		&  &best-fit  &$mean\pm\sigma$&best-fit  &$mean\pm\sigma$\\\hline
		&{\boldmath$\Omega{}_{m }  $} &$0.3113$ &$0.3148\pm0.0067$ &$0.2903$ &$0.2888\pm0.0044$\\

		&{\boldmath$\log_{10}\alpha$} &$7.59$&$7.88^{+0.15}_{-0.12}$ &$7.72$ &$7.14^{+0.61}_{-0.34}$\\
		
		&{\boldmath$k_{0}$} &$1.68$&$1.70^{+0.19}_{-0.12}$ &$0.988$ &$ 1.003^{+0.049} _{-0.065}$\\
		
		$k$-DM(1)&{\boldmath$c_g^2$} &$0.052$&$<0.014$ &$0.0116(10^{-4})$ &$<0.0373(10^{-4})$\\
		
		&{\boldmath$w$} &$0.011$ &$<0.009$&$0.181(10^{-4})$ &$0.126\pm0.049(10^{-4})$ \\
		
		&{\boldmath$H_0$}  & $67.67$&$67.40\pm0.49$ &$69.25$ &$69.36\pm 0.36$ \\
		
		&{\boldmath$S_8            $}  & $0.8280$&$0.829\pm0.012$ &$0.7651$ &$0.7649^{+0.0067}_{-0.0083}$\\
		\hline
		&{\boldmath$\Omega{}_{m }  $} &$0.3107$ &$0.3129\pm0.0073$ &$0.2954$ &$0.2969^{+0.0054}_{-0.0066}$\\

		&{\boldmath$\log_{10}\alpha         $} &$7.65$ &$7.7^{+1.4}_{-1.1}$ &$6.54$ &$6.49^{+0.77}_{-0.63}$\\

		$k$-DM(2)		&{\boldmath$k_{0}          $} &$2.51$ &$12^{+15}_{-10}$ &$1.66$ &$ 1.93^{+0.11}_{-0.30}$\\

		&{\boldmath$w              $} &$0.570$ &$<0.532$ &$1.18(10^{-7})$ &$<5.48(10^{-5})$ \\

		&{\boldmath$H_0            $}  &$67.66$ &$67.54\pm0.53$ &$68.84$ &$68.67^{+0.55}_{-0.48}$ \\

		&{\boldmath$S_8            $}  &$0.822$ &$0.826\pm0.013$ &$0.7795$ &$0.7875\pm0.0099$\\\hline

		&{\boldmath$\Omega{}_{m }  $} &$0.313328$ & $0.3142\pm0.0065$ &$0.2929$ &$0.2933\pm0.0045$ \\

		CDM &{\boldmath$H_0$} &$67.52$ &$67.46\pm0.47$ &$69.47$ &$69.02\pm0.37$ \\
		
		&{\boldmath$S_8$} & $0.828789$ &$0.831\pm0.0012$ &$0.7872$ &$0.7833\pm0.0074$ \\\hline

\end{tabular}} \label{tab:best}    

\end{table*} 
However, the CDM fluid velocity in the conformal Newtonian gauge is not zero in general. In  $k$-space, equation~\ref{eq-wdm-n} gives
\begin{equation}
	\dot{\delta}_{\rm CDM}=-\theta_{\rm CDM}+3\dot{\phi}, \qquad \dot{\theta}_{\rm CDM}=-\frac{\dot{a}}{a}\theta_{\rm CDM}+k^2\psi.
\end{equation}
As we mentioned, in this study, we intend to consider dark matter such that its behavior changes in terms of
scale, so it behaves as relativistic such as warm dark matter particles in large $k$ scales, and as non-
relativistic such as cold dark matter in small $k$ scales.
Therefore, we introduce a step function, ${\cal{S}}(k)$, for switching between these two boundary conditions.
${\cal{S}}(k)$ could be any kind of step (switching) function; for example we consider it as 
\begin{equation}
	{\cal{S}}(k)=\frac{1+\tanh[\alpha(k-k_0)]}{2}.
\end{equation}
$ \alpha $ and $k_0$ are free parameters that $\alpha$ control the smoothness of the transition between cold and warm dark matter. We rewrite~\ref{eq:wdm} by using ${\cal{S}}(k)$ as
\begin{equation}\label{eq-wdm-s}
	\dot{\delta}=\frac{\dot{h}}{2}+{\cal{S}}(k)\big[-3{\cal{H}}(c_s^2-w)\delta-(1+w)\theta\big],
\end{equation}
\begin{equation}\label{eq-wdm-s2}
	\dot{\theta}={\cal{S}}(k)\big[-{\cal{H}}(1-3c_g^2)\theta+\frac{c_s^2}{1+w}k^2\delta-k^2\sigma\big],
\end{equation}
\begin{equation}\label{eq-wdm-s3}
	\dot{\sigma}={\cal{S}}(k)\bigg[-3\big(\frac{1}{\tau}+{\cal{H}}(\frac{2}{3}-c^2_g-\frac{1}{3}\frac{{\cal{P}}}{p})\big)\sigma+\frac{4}{3}\frac{c^2_{\rm vis}}{1+w}(2\theta+\dot{h})\bigg].	 
\end{equation}
It is obvious that if ${\cal{S}}(k)$ vanishes, the above equations reduce to the CDM. This case happens for $k<k_0$ and is more precise for larger $\alpha$'s.\\
We implement the above equations in the public Boltzmann solver \texttt{CLASS}. Since we expect this model to behave similarly to CDM in the cosmological background, we only change the perturbation equations in module \texttt{perturbation.c}. 
We analyzed this model in two cases:  i) the sound speed behaves independently of $k$ as a constant parameter ($k$-DM(1)) and ii) the case where it changes depending on $k$ given to equation~\ref{cs} ($k$-DM(2)).\\
\section{$\lowercase{k}$-Dependent DM model verse data}\label{sec:mcmc}
In this section, we present constraints on the $k$-Dependent dark matter model we have introduced. For MCMC analysis, we use the Metropolis-Hastings algorithm  of the cosmological sampling package \texttt{MONTEPYTHON-v3}, connected to an altered version of the Boltzmann Solver \texttt{CLASS}.\\
We use the following dataset combination to perform statistical inference:
\begin{figure*}
	\includegraphics[width=8cm, height=7cm]{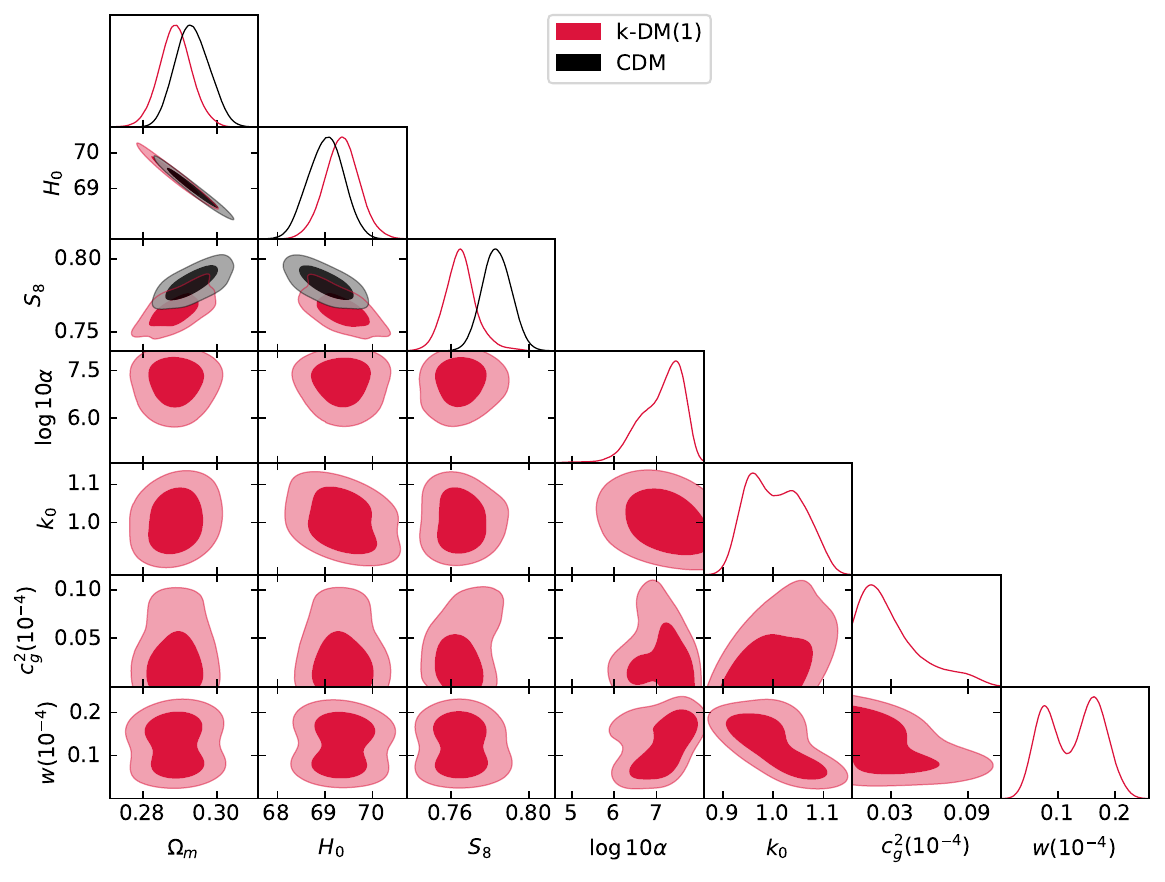}
	\includegraphics[width=8cm, height=7cm]{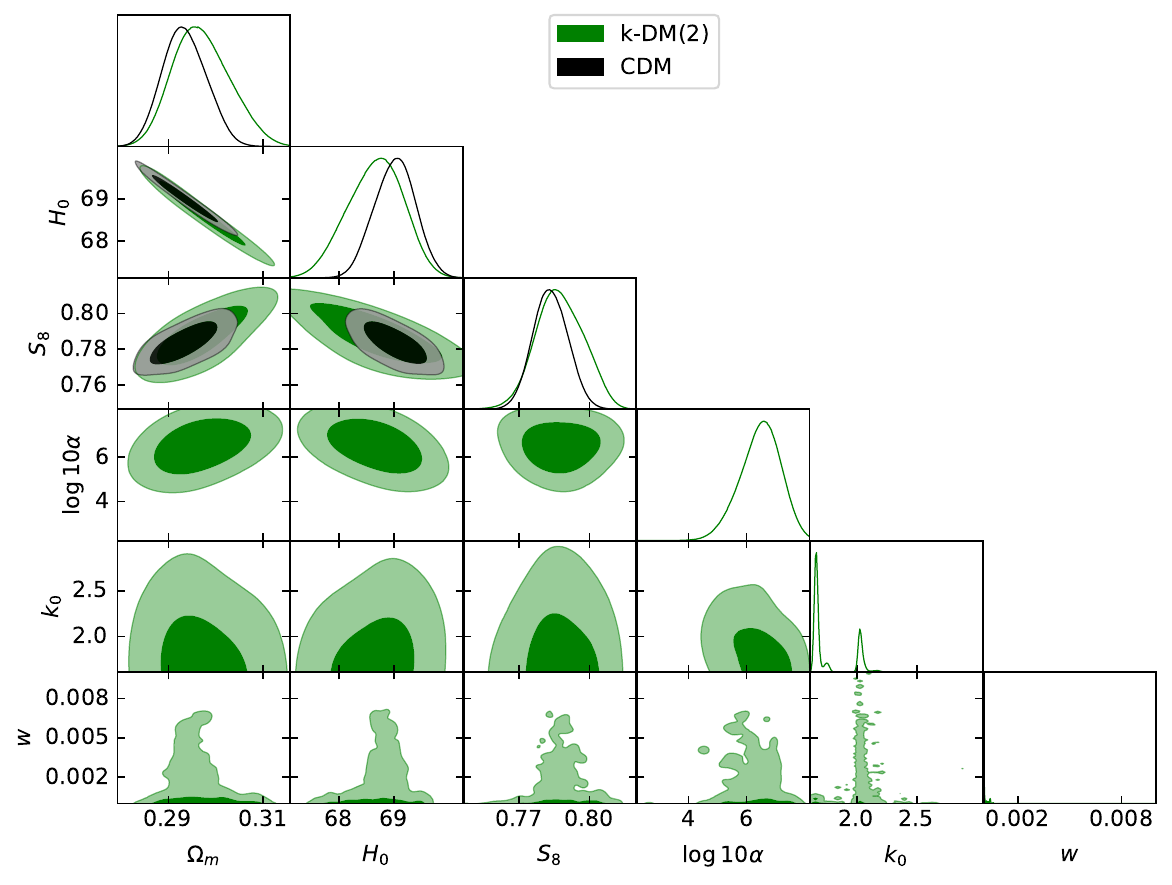}
	\caption{1D likelihoods and 2D contours in 68\% and 95\% CL marginalized joint regions for		
		 chosen free parameters while they are constrained by using $Planck+Other$ datasets. It seems the $k$-DM models predict lower $S_8$ to alleviate this tension while the $H_0$ value is not affected too much.
		\label{fig1}}
\end{figure*}
\begin{itemize}
	\item CMB:
	We use the CMB temperature and polarization auto- and cross-correlation measurements of the most recent Planck 2018 legacy release, including the full
	temperature power spectrum at multipoles 2 $\leq$ l $\leq$ 2500 and the polarization
	power spectra in the range 2 $\leq$ l $\leq$ 29 (lowP). We also include information on the gravitational lensing power spectrum estimated from the CMB trispectrum analysis. \citep{Aghanim:2018eyx}.
	\item  BAO:
	We use the BAO measurements from the Baryon Oscillation Spectroscopic Survey Data Release 12 (BOSS DR12)~\citep{BOSS:2016wmc} ,SS DR14-Ly-$\alpha$ combined correlations~\citep{deSainteAgathe:2019voe}, Lyman-$\alpha$
	forest autocorrelation~\cite{deSainteAgathe:2019voe}, and the cross correlation of Lyman-$\alpha$ and QSO~\citep{Blomqvist:2019rah}.
	\item LSS: We use three  different sets of LSS data in order to check whether $k$- Dependent dark matter model leads to a suppression in the matter power spectrum relative to the CDM:\\
	\textbf{1-} KiDS + Viking 450 (KV450) matter power spectrum shape data; this combined analysis of data from the KiloDegree Survey (KiDS) and the VISTA Kilo-Degree Infrared Galaxy Survey (VIKING) includes photometric redshift measurements with cosmic shear/weak-lensing observations to measure the matter power spectrum over a wide range of $k$-scales at redshifts between 0.1 and 1.2~\citep{Hildebrandt:2018yau}.\\
	\textbf{2-} Planck SZ (2013):  Another independent LSS dataset is the Planck SZ which studies the properties of galaxy clusters by measuring the Sunyaev-Zeldovich effect. But we should note that the measurements of galaxy distribution from the SZ effect depend on a mass bias factor $ (1-b) $ that relates the observed SZ signal to the true mass of galaxy clusters. In Planck SZ (2013), a numerical simulation of the $ S_8 $ measurement is reported by fixing the mass bias to its central value $(1-b) = 0.8$.
	Later,  the Planck SZ (2015) report allowed $ (1-b) $ to vary with a Gaussian prior centered at 0.79. The central value of the resulting $ S_8^{\rm SZ}$ becomes smaller but has a much larger uncertainty,
	$ S_8^{\rm SZ}= 0.744\pm0.034$, and less tension to CMB	measurements~\citep{Planck:2015lwi}. For our analysis, we chose this data set since the central value $\sigma_8$ of the SZ (2013) analysis is consistent with many low-redshift measurements~\citep{Zu:2023rmc}.\\
	\textbf{3-}WiggleZ $P(k)$ data: Since dark energy has an effect on the expansion history of the Universe and on the growth of cosmological structures, we also use WiggleZ data in this study.
	The WiggleZ Dark Energy Survey is a survey	to measure the large scale structure of the Universe by mapping the distance-redshift relation with baryon acoustic oscillations~\citep{Kazin:2014qga}.
\end{itemize}
We employ the $\chi^2$ statistics to constrain our theoretical model as:
\begin{equation}
	\chi^2=\frac{({\cal{P}}_{\rm obs}-{\cal{P}}_{\rm th})^2}{\sigma_{{\cal{P}}}^2},
\end{equation}
here ${\cal{P}}_{\rm obs}$, ${\cal{P}}_{\rm th}$ and $\sigma_{{\cal{P}}}^2$ indicate the observed values, the
predicted values and the standard deviation, respectively. Note that in addition to the six free parameters of the standard model, i.e, $(\Omega_b,\Omega_{DM},10\theta_{\rm MC},\ln 10^{10}A_s,n_s,\tau_{\rm reio})$, the $k$-Dependent dark matter model introduced in the previous section includes for the first case: $(\alpha, k_0, w,c^2_g)$ and the second case: $(\alpha, k_0, w)$. To span the $\alpha$ parameter's space, we work with  $ \log_{10}\alpha $ instead of $\alpha$. The flat priors we assumed for the parameters are given by $k_0\in[0,{10}]$, $\alpha\in[0,10^{10}]$,  $w\in[0,1]$, and $c^2_g\in[0,1]$. 
The convergence of chains for each parameter is measured by the Gelman-Rubin criterion, and one can obtain acceptable $ R-1 $ values (i.e., below 0.01 for every parameter) with an iterative strategy~\citep{Gelman:1992zz} and the average acceptance rate (acc) is around 0.2. \\
In order to check the cosmic tensions in these models, we added data step by step in two MCMC scans
as: Planck and then Planck+Other. This can provide us with further intuition as a starting point, given that
Planck’s data has provided the most precise measurements of the early universe.\\
In the Table~\ref{tab:best}, we report the best and the mean values and 68\% CL intervals for the main parameters, including the total matter density
parameter ($\Omega_m=\Omega_B+\Omega_{DM}$), the present-day
expansion rate of the Universe or the Hubble constant, $H_0$, and $S_8=\sigma_8\sqrt{\Omega_m/0.3}$ in different scenarios for two MCMC analyzes. We also show posterior distributions (1$\sigma$ and 2$\sigma$ intervals) as dark and light-shaded contours for  MCMC analysis, respectively, in the plots of the Figure~\ref{fig1}.\\
Some points in these plots need to be stressed. First, it is clear that by considering all different data sets, the
parameters are bound more tightly than the analysis with Planck data\footnote{The only Planck-constrained parameters are not shown in the figures, but it has checked that Planck and Planck+Other are consistent. This means the contours for the latter are inside the only Planck contours.}. Second, we can see that the decrease of $ H_0 $ values are associated with the increase of $ \Omega_m $ values and vice versa in both $k$-Dependent and cold dark matter scenarios. 
\begin{figure*}
	
	\includegraphics[width=10cm]{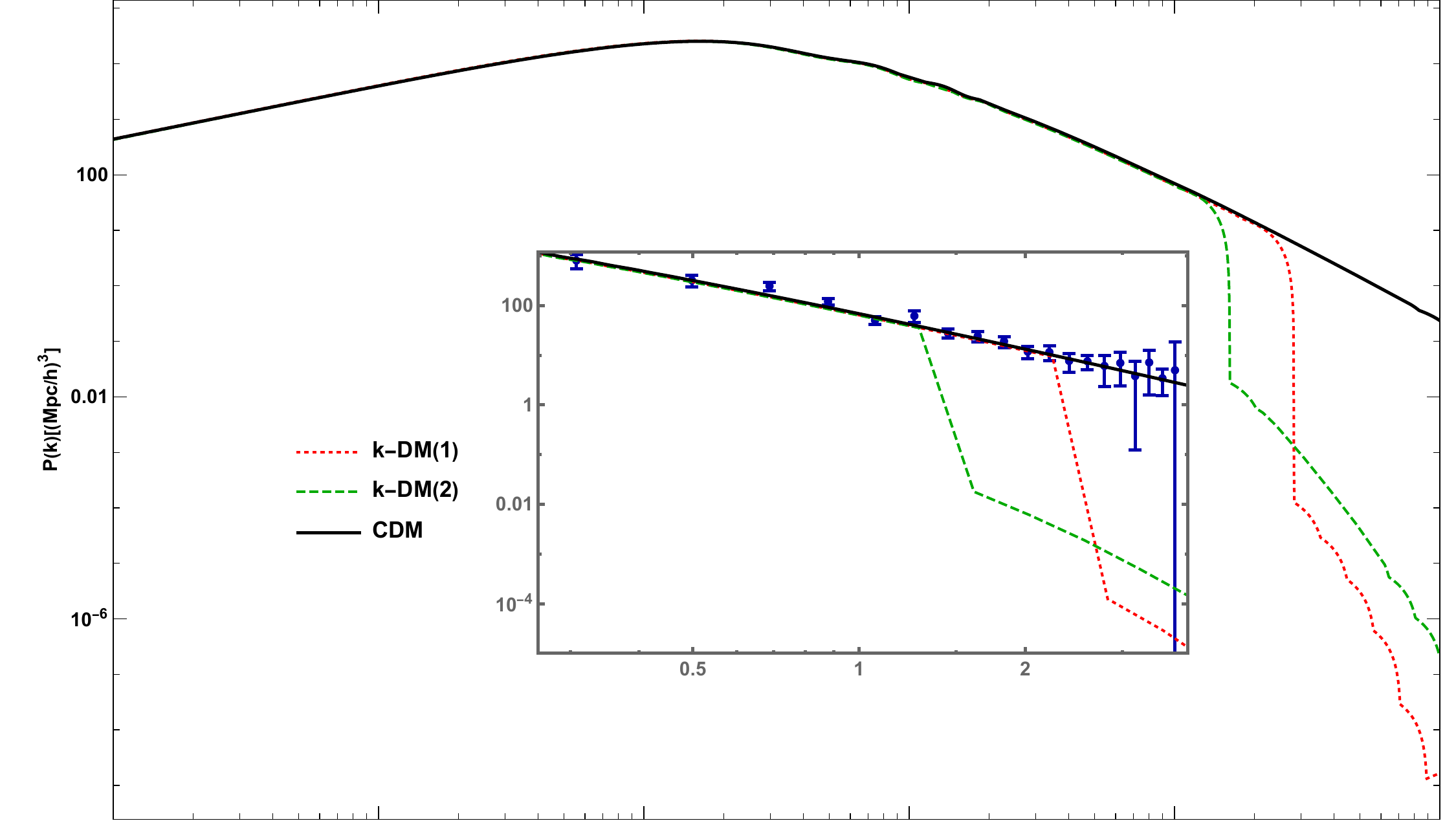}
	\includegraphics[width=10cm]{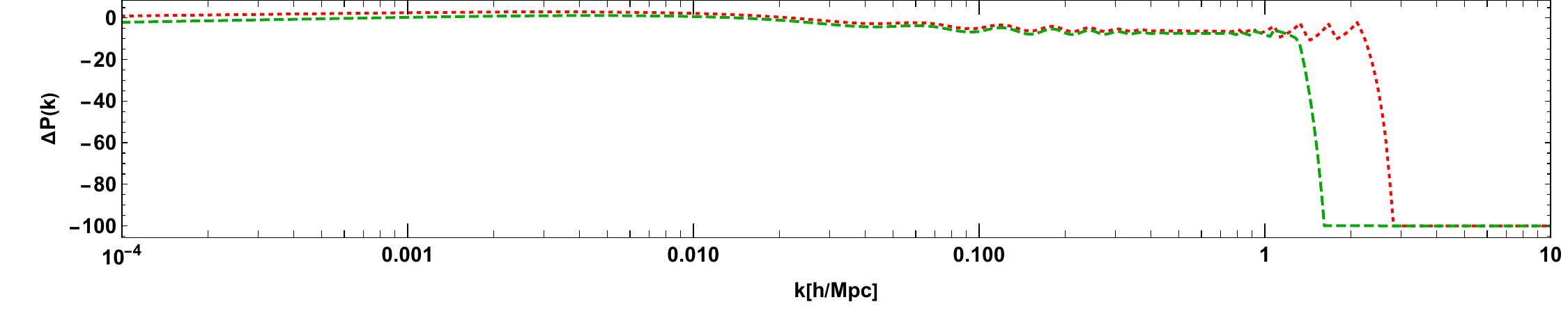}
	\caption{The matter power spectrum for $k$-Dependent DM and CDM models, and the fractional difference between them. The behavior of $k$-DM models mimics the CDM for small $k$'s.
		However, we see a transition for large $k$'s in $k$-DM models. However, there is no very precise data points at those scales.
	}
	\label{figpk}
\end{figure*}
As we see in Table~\ref{tab:best}, assuming $k$ dependence of dark matter behavior for the Planck+Other analysis seems to improve the $ S_8 $ tension for the $k$-Dependent dark matter scenario. However, we do not see any significant improvement in addressing the $H_0$ tension. Note that the results show a small deviation from CDM due to non-zero values for $w$ and $c_g^2$ when we have Planck+Other datasets. Their values are at order $ {\cal{O}} (10^{-5}) $, which are in agreement with generalized dark matter models~\citep{Ilic:2020onu,Kopp:2016mhm}.

\begin{figure}
	\begin{center}
		\includegraphics[width=8.5cm]{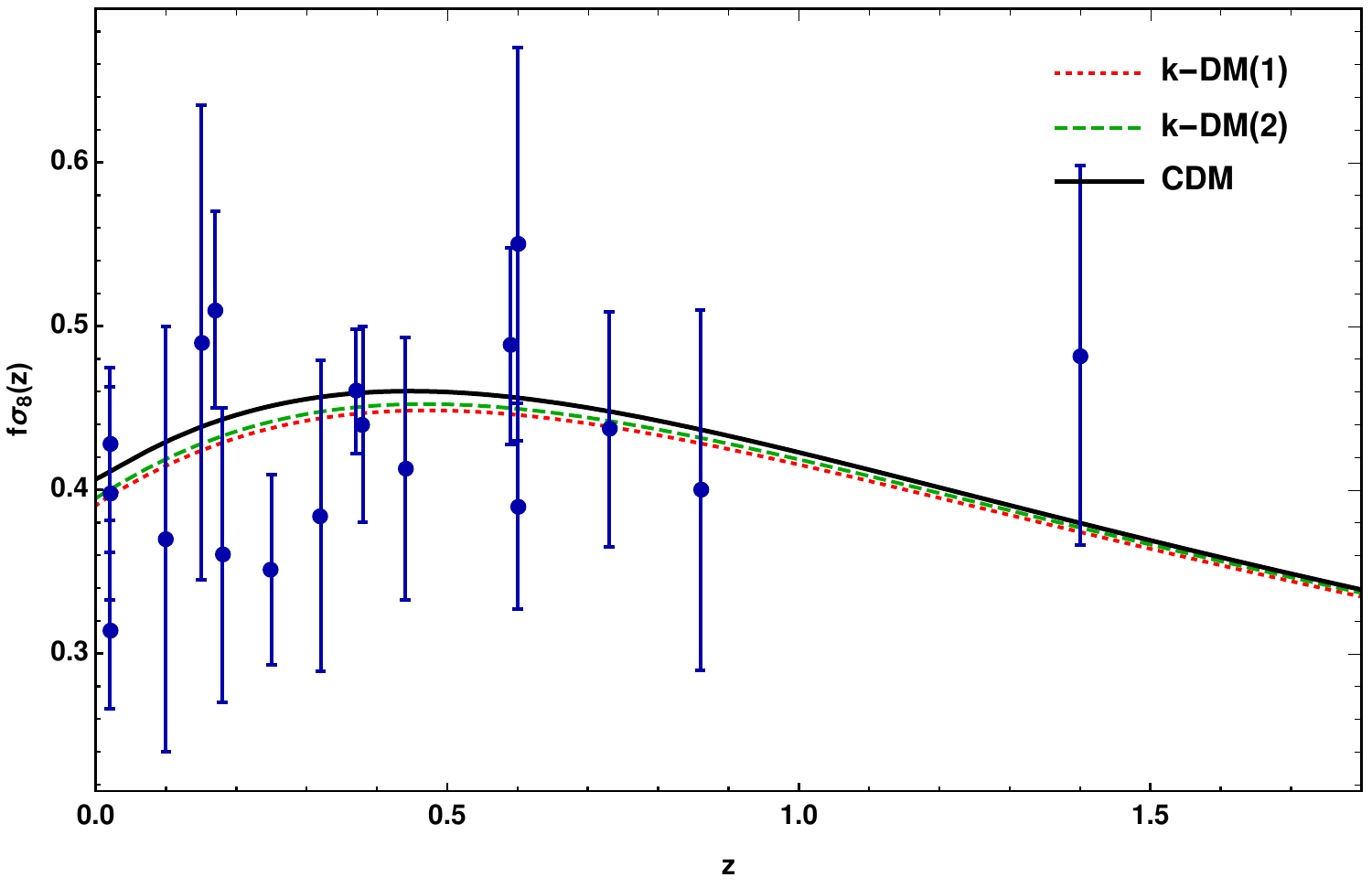}
		\caption{The growth rate of matter fluctuations for $k$-Dependent DM model compared to CDM
			model. The observational constraints
			are taken from \citep{Kazantzidis:2018rnb}.
		}
		\label{figfs8}
	\end{center}
\end{figure}

Next, to check whether the fit is good and also to choose the best and most compatible model with the observational data, we employ the simplest method that is usually used in cosmology, which is called the least squares method,  $\chi^2_{{}_{\rm tot}}$. In this case, the model with smaller $\chi^2_{{}_{\rm tot}}$ is taken to be a better fit to the data~\citep{Davari:2021mge}. Comparing $k$-Dependent model to the CDM scenario, we note that $k$-Dependent DM model does better than the CDM model. 
However, one can have the impression that the model with the lowest  $\chi^2_{{}_{\rm tot}}$ is not necessarily the best because adding more flexibility with extra parameters will normally lead to a lower  $\chi^2_{{}_{\rm tot}}$.  In this work, the $k$-Dependent model has three more parameters than the CDM scenario. In order to deal with model selection, a standard approach is to compute the Akaike Information Criterion (AIC). It is defined as
\begin{equation}\label{aic}
	AIC=\chi^2_{\rm min}+2M+\frac{2M(M+1)}{N-M+1},
\end{equation}
where $M$ is the number of free parameters in the model and $N$ is the number of data points; thus, $\Delta AIC=\Delta \chi^2_{\rm min}+2\Delta M$. We neglect the third term in the Equation~\ref{aic} for large sample sizes, $N\geq M$. We report the result of MCMC analysis for the best-fit $\chi^2_{min}$ for observational Planck and total data sets and for both models in Table~\ref{tab:com}.\\
The results of this analysis can be interpreted with the Jeffreys' scale as follows: among all models, the one that minimizes the AIC is considered to be the best one, and
if the difference between the AIC of a given model and the best model is
smaller than 4, one concludes that the data equally support the best fitted model and a given model. In the case of $ 4 <|\Delta AIC| < 10 $, observations still support the given model but less than the
best one. Finally, for $|\Delta AIC| > 10 $, observations basically do not support the given model compared to the best model~\citep{Davari:2018wnx}. According to Table \ref{tab:com}, the only Planck data prefers CDM with respect to $k$-DM models. However, adding the other datasets make the situation in favor of $k$-DM models. This may mean that $k$-DM models have more space to include all the datasets altogether.
\begin{table*}
	\caption{The result of MCMC analysis for the best-fit $\chi^2$, and AIC. It shows that $k$-DM models have more space to have Planck+Other datasets altogether consistently.} 
	\centering 
	\begin{tabular}{|c|c|c|c|c|c|c|} 
		\hline   
		Parameters &\multicolumn{2}{c|}{CDM}  &\multicolumn{2}{c|}{k-DM(1)}&\multicolumn{2}{c|}{k-DM(2)}  
		\\\cline{2-7}   
		&$Planck$  &$Planck+Other$ &$Planck$  &$Planck+Other$&$Planck$  &$Planck+Other$
		\\  \hline
		{\boldmath$\chi^2_{\rm min}$}  &$2780.9$&$3824$  &$2781.02$&$3812.2$&$2780.44$&$3816.7$  \\\hline
		{	\boldmath$\rm AIC_{k-DM(i)}-AIC_{CDM}$ }&$0$ &$0$ &$8.12$ &$-3.8$&$5.54$ &$-1.3$\\
		\hline		
	\end{tabular} \label{tab:com}
\end{table*}
In order to have a better understanding of the aspects of obtaining from MCMC scans, in the following, we discuss the features of the $k$-Dependent model in the CMB and the matter power spectrum.
In Figure~\ref{figpk}, we show the matter power spectrum, $  P(k) \equiv \langle\delta_m(k)\rangle ^2$, in the $k$-Dependent model
relative to the CDM model for the best obtained values using Planck+Other data. As we see, $k$-Dependent dark matter case mimics the CDM scenario to $k\simeq1.3$ for the $k$-DM(1) and $k\simeq2.2h Mpc^{-1}$ for the $k$-DM(2), but then starts to deviate
at larger $k$'s (i.e., small scales) and suppresses the power spectrum of matter by a large difference compared to the standard model. We include the information embedded in the Ly$\alpha$ forest measured with the eBOSS-DR14 data release on scales of a few Mpc. One reason for this difference could be the lack of observational data in this range. In Figure~\ref{figfs8}, we notice that considering $k$ dependence for dark matter has the influence of slowing down the evolution rate of the dark matter perturbations. This means that structures cluster slower, as we predicted from the Figure~\ref{figpk} for $k>1$ with a slight difference $ P_{k-DM(i)}(k)<P_{CDM}(k) $.\\
\begin{figure}
	\begin{center}
		\includegraphics[width=8.5cm]{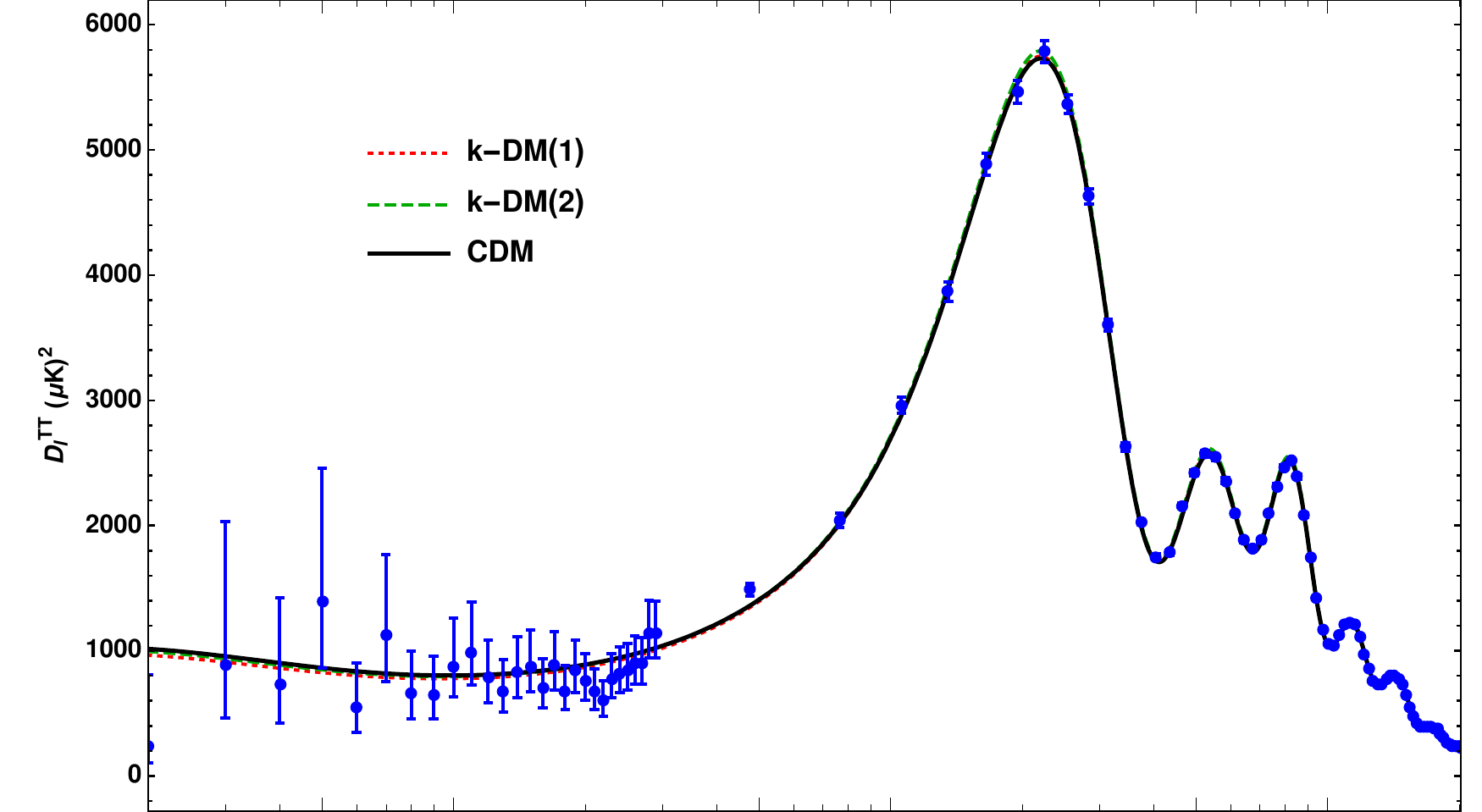}
		\includegraphics[width=8.5cm]{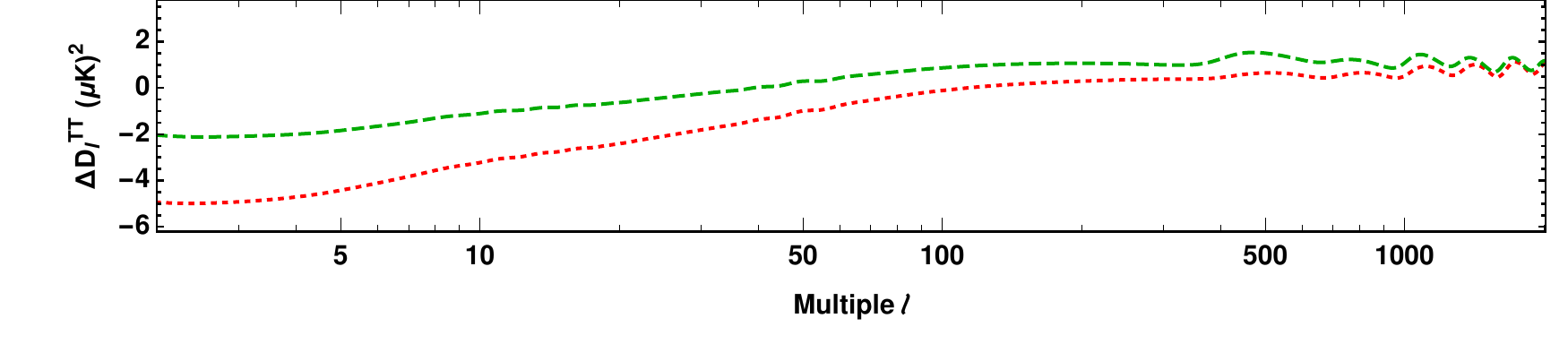}
		\caption{ Temperature anisotropies in the CMB. The bottom part of the panel displays the relative temperature differences between the $k$-DM and the CDM model.
		}
		\label{figtt}
	\end{center}
\end{figure}
Since the Planck collaboration has measured the temperature and polarization maps of the CMB very precisely, it has placed stringent limits on the parameter space of the CDM model. This motivates us to study $k$-Dependent DM signatures in CMB maps. In Figure~\ref{figtt}, we show how $k$ dependency affects
the temperature power spectra, including the variation with respect to the CDM model. We can see a suppression in the amplitude of the lower multipoles in the temperature power spectrum. As we know, the integrated Sachs-Wolfe (ISW) effect is important on such scales. Also, the small $l$'s of CMB (TT and, even better, EE) give information on the reionization history. We obtain the redshift of the reionization, $ z_{\rm reio} $, using the best values of parameters to be 6.02 , 6.41 and 7.38 for $k$-DM(1), $k$-DM(2) and CDM respectively. The $z_{\rm reio}$ of the $k$-DM(1) model has the biggest differences from the standard model. 

\begin{figure}
	\begin{center}
		\includegraphics[width=8.5cm]{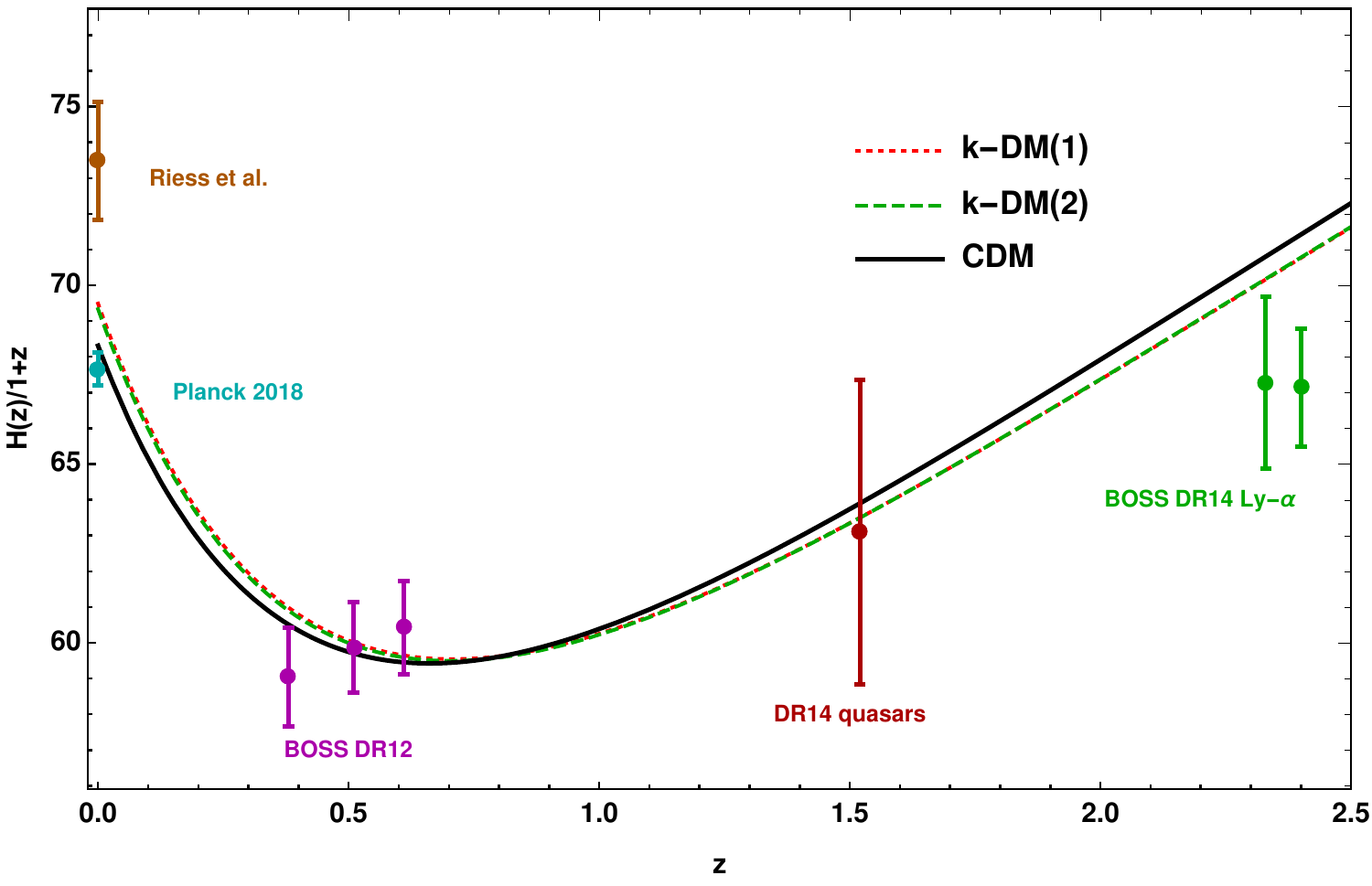}
		\caption{The redshift evolution of H(z)/1+z using the best fit values of cosmological parameters in Tables~\ref{tab:best} by Planck+Other data for the proposed $k$-DM scenarios and CDM model. Also, in the Figure, the observational data points from the Riess et al. ~\citep{Riess:2019cxk}, Planck 2018~\citep{Aghanim:2018eyx}, BOSS DR12~\citep{BOSS:2016wmc},
			DR14 quasars~\citep{Zarrouk:2018vwy}, and BOSS DR14 Ly-$\alpha$ ~\citep{deSainteAgathe:2019voe} measurements have
			been specified for comparison. 
				}
		\label{fig:h}
	\end{center}
\end{figure}

\begin{figure}
	\begin{center}
		\includegraphics[width=4cm]{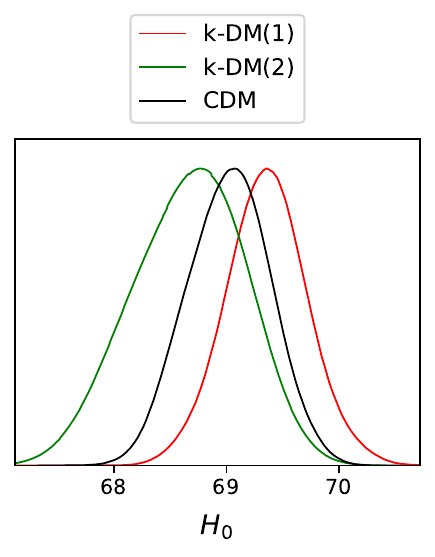}
		\includegraphics[width=4cm]{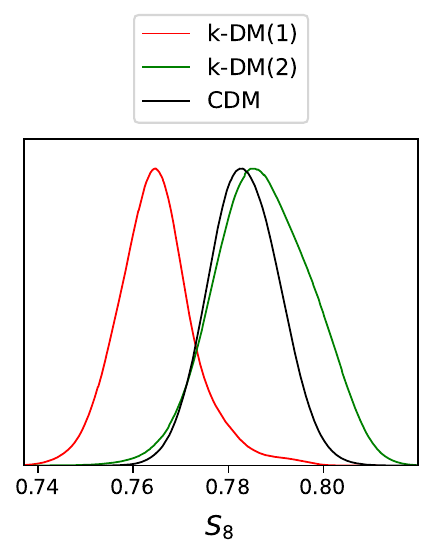}
		\caption{Comparison between $H_0$ and $ S_8 $ tensions for all considered models in this study using the Planck+Other data.
		}
		\label{figten}
	\end{center}
\end{figure}

A crucial quantity in determining the age and evolution of the universe in cosmology is $ H_0 $. It represents the current rate of expansion of the universe.
Because of the impact of Hubble’s expansion on the growth of matter perturbations, it is significant
to survey the behavior of H(z) in various DM cosmologies. We plot the evolution of H(z)/1+z in Figure~\ref{fig:h}.  
Our results in Tables~\ref{tab:best} and Figure~\ref{figten} show that the assumption of dependence dark matter to $k$ scale can only reduce the $ S_8 $ tension and not the $ H_0 $ tension. In general, $k$-DM(1) model, which considered the equation of state, $w$, and adiabatic sound speed, $c^2_g$ independent of $k$ scale, reduces $ S_8 $ tension more than other models.
\section{Discussion}
The warm dark matter model has always been of interest mainly because of the possible need to alleviate the small-scale problems of the $\Lambda$CDM. With such insight and also motivated by the effect of adding this cosmological component to reduce current cosmological tensions, in this work we considered a scenario in which the behavior of dark matter depends on the scale. It mimics CDM for small $k$'s and WDM for large $k$'s. A motivation for us was to check if the trace of WDM, which can be seen in very small scales to address e.g., can the core-cusp problem show itself in the (very short) cosmological scales? Our results show that this transition can affect the amplitude of the matter fluctuations, such that reducing the $S_8$ tension. However, it seems the lack of cosmological data for very large $k$'s makes it hard to answer to the above question. For future analysis, we can think of a more theoretical framework and also find cosmological datas at very small scales which are cleaned from the baryonic physics. One way can be tracing the effects of our model in non-linear structure formation and the dark matter halo distributions.
\section{ACKNOWLEDGMENTS}
NK would like to thank Marzieh Farhang for instructive discussions during working on~\cite{Farhang:2023hen}. This work has been supported financially by a grant from Basic Sciences Research Fund under grant number BSRF-phys-399-06.
ZD also acknowledges support from Iran Science Elites Federation under grant number M401543.
\section{DATA AVAILABILITY}
No new data were generated or analysed in support of this research.
\bibliography{refnew}{}
\bibliographystyle{mnras}
\end{document}